\documentstyle[14pt]{article}

\begin{document}           
\title{
 Report IRB-StP-GR-120696 \newline
\newline \newline \newline
{\bf
The Analysis of Initial Conditions for the LTB Model
}
}
\author{\it by \\
\\
{\bf Alexander Gromov}
\\ \\
\small\it St. Petersburg State Technical University \\
\small\it Faculty of Technical Cybernetics, Dept. of Computer Science \\
\small\it 29, Polytechnicheskaya str. St.-Petersburg, 195251, Russia \\
\small and \\
\small\it Istituto per la Ricerca di Base \\
\small\it Castello Principe Pignatelli del Comune di Monteroduni \\
\small\it I-86075 Monteroduni(IS), Molise, Italia \\
\small\it e-mail: gromov@natus.stud.pu.ru
}
\date{}
\maketitle
\begin{abstract}
The LTB model is studied as the Caushy problem for the equations defined
two metrical functions $\lambda(\mu,\tau)$ and $\omega(\mu,\tau)$.
The initial conditions throught metrical functions are presented.
The rules of calculating three undetermined functions
$f(\mu)$, $F(\mu)$ and ${\bf F}(\mu)$ are obtained. The general expressions
for the density and Habble function in the LTB model are written by the
metrical functions.
\\ \\
PACS number(s):98.80

\end{abstract}
\newpage

\section{The Introduction} \label{introd}

The LTB model is one of the most known spherical symmetry model in general
relativity. It was created by
Lemaitre \cite{Lemaitre:33}, Tolman \cite{Tolman:34}, and Bondy
\cite{Bondy:47}
during the period of time from 1933 to 1947. The exact
solution have been obtained by Bonnor \cite{Bonnor:72} in 1972 and
\cite{Bonnor:74} in 1974.
The LTB model represented one of the simplest nonhomogeneous nonstationary
cosmological models and due to this fact
is used to study some new ideas in the cosmology.

The present interest to LTB was risen by the observations
shown the fractal structure of the Universe in the
large scale \cite{Coleman}, \cite{Pietronero:87}, and discass in the set of
article (the modern review is presented in \cite{Bar}).
In a set of papers the LTB model is used to study the
observational datas and
redshift as a main cosmological test \cite{Ribeiro:92a} -
\cite{Moffat:94b}.

The central problem of
using the LTB model is in calculation of three undetermined
functions which defined the solution. There is a set of ways how to solve
this problem from the physical point of view in the mentioned papers.
This article is devoted to the mathematical point of view on this matter
\cite{A.G.}.

\section{The LTB Model} \label{T_sol}

This section is devoted to presentation the Lemaitre-Tolman-Bondy model
and section {2} of the paper \cite{Tolman:34} is cited.
The co-moving system of coordinate is used in this model where the interval
has the form
\begin{equation}
ds^2(r,t) = -e^{\lambda(r,t)}dr^2 - e^{\omega(r,t)}
\left(
d\theta^2 + sin^2\theta d\phi
\right) + dt^2.
\label{ds2}
\end{equation}
$\lambda(r,t)$ and $\omega(r,t)$ are metrical functions \cite{Tolman:34}
definding the solution. Following the \cite{Tolman:34} we will omit the
arguments in the functions $\omega$ and $\lambda$ in this section.
In the co-moving system of coordinate with the line element (\ref{ds2})
the energy-momentum tensor
\begin{eqnarray}
T^{\alpha,\beta} = \rho \frac{dx^{\alpha}}{ds}\frac{dx^{\beta}}{ds}
\nonumber
\end{eqnarray}
has only one non zero component
\begin{displaymath}
T^{4}_{4} = \rho \qquad
T^{\alpha}_{\beta} = 0, \qquad \alpha \quad \mbox{or} \quad \beta =
4.
\label{T_ik:RO}
\end{displaymath}
Using them together with Dingle results \cite{Dingle} we obtain the system
of equations of the LTB model:
\begin{equation}
8 \pi T^{1}_{1} = e^{-\omega} -
e^{-\lambda}\frac{{\omega^{\prime}}^2}{4} + \ddot \omega +
\frac{3}{4} \dot \omega^2 - \Lambda = 0
\label{T:4}
\end{equation}
\begin{eqnarray}
8 \pi T^{2}_{2} = 8 \pi T^{3}_{3} = \nonumber\\
- e^{- \lambda}
\left(
\frac{\omega^{\prime\prime}}{2} +
\frac{\omega^{\prime}{}^{2}}{4} -
\frac{\lambda^{\prime}\omega^{\prime}}{}
\right)
+\frac{\ddot \lambda}{4}
+\frac{\dot\lambda^2}{4}
+ \frac{\ddot \omega}{2}
+\frac{\dot \omega^2}{4}
+\frac{\dot \lambda \dot \omega}{4}
- \Lambda = 0
\label{T:5}
\end{eqnarray}
\begin{equation}
8 \pi T^{4}_{4} =e^{-\omega} - e^{-\lambda}
\left(
\ddot \omega +\frac{3}{4}\dot \omega^2 - \frac{\lambda^{\prime}
\omega^{\prime}}{2}
\right)
+ \frac{\dot \omega^2}{2}
+ \frac{\dot \lambda \dot \omega}{2}
- \Lambda = 8 \pi \rho
\label{T:6}
\end{equation}
\begin{equation}
8 \pi e^{\lambda} T^{1}_{4} = - 8 \pi T^{4}_{1} =
\frac{\omega^{\prime}\dot \omega}{2}
- \frac{\dot \lambda \omega^{\prime}}{2}
+\dot \omega^{\prime} = 0,
\label{T:7}
\end{equation}
where
\begin{eqnarray}
{}^{\prime} = \frac{\partial}{\partial r} \qquad
\dot{} = \frac{\partial}{\partial t}
\nonumber
\end{eqnarray}
The equation (\ref{T:7}) has the solution
\begin{equation}
e^{\lambda} = e^{\omega} \frac{\omega^{\prime \,2}}{4 f^2(r)}
,
\label{T:8}
\end{equation}
where $f(r)$ is undetermined function.
Substituting (\ref{T:8}) into (\ref{T:4}) we obtain
\begin{equation}
e^{\omega}
\left(
\ddot \omega +\frac{3}{4} \dot \omega^2 - \Lambda
\right)
+
\left[
1 - f^2(r)
\right] = 0.
\label{T:10}
\end{equation}
This equation is integrated twice.
First integral gives the equation
\begin{equation}
e^{3 \,\omega / 2}
\left(
\frac{\dot \omega^2}{2} - \frac{2}{3} \Lambda
\right)
+2 e^{\omega /2} \left[1 - f^2(r)\right] = F(r),
\label{T:11}
\end{equation}
and the second one gives the equation
\begin{equation}
\int
\frac
{{\rm d e^{\omega / 2}}}
{
\sqrt{{\rm
f^2(r) - 1 +\frac{1}{2} } F {\rm (r) e^{-\omega/2} + \frac{\Lambda}{3}
e^{\omega}}
}
}
 = t +  {\bf F} {\rm (r)
}
\label{T:12}
\end{equation}
The equations (\ref{T:11}) and (\ref{T:12}) hold undetermined functions
$\it F(r)$ и $\bf F\it(r)$.
The substitution of (\ref{T:8}) into (\ref{T:6}) together with
(\ref{T:11}) gives the equation for density
\begin{equation}
8 \pi \rho = \frac{1}{\omega^{\prime}e^{3\omega/2}}
\frac{
\partial
F{\rm (r)}}{{\rm \partial r}}
\label{T:15}
\end{equation}

\section{The Cauchy Problem for the LTB \newline Model}

Before we study the LTB model let us introduce the follow
characteris\-tic values:
a velocity of light $c$, an observational meaning of the Habble
constant $\tilde H$, a characteristic time $1/\tilde H$ and characteristic
length $c/ \tilde H$.
We use the co-moving system of coordinates in the LTB model, so the radial
coordinate $r$ has the sense of Lagrangian mass coordinate \cite{LE&AD},
\cite{L&L}.
Two dimensionless variables $\mu$ and $\tau$ are definded by the rules
\begin{eqnarray}
\mu = \frac{r}{m} \quad \quad \quad
\tau = \tilde H t,
\nonumber
\end{eqnarray}
where $m$ is a full mass of the "gas".
The dimensionless Habble function $h(\mu,\tau)$ and density
$\delta(\mu,\tau)$ will be also used:
\begin{eqnarray}
h(\mu,\tau) = \frac{H(\mu,\tau)}{\tilde H}, \qquad
\delta(\mu,\tau) = \frac{\rho(r,t)}{\rho(0,0)},
\nonumber
\end{eqnarray}
where $H(0,0) = \tilde{H}$.

Let us write the interval (\ref{ds2}) as
\begin{equation}
ds^2(r,t) = -A e^{\lambda(r,t)}dr^2 - B e^{\omega(r,t)}
\left(
d\theta^2 + sin^2\theta d\phi
\right) +c^2dt^2,
\label{ds2_1}
\end{equation}
where two constants $A$ and $B$
are introduced to take into account
the fact that (\ref{ds2_1}) is dimension equation.

The dimension of $\left[ ds^2 \right]$ is $L^2$, dimension of
$\left[ A \right]$ is $L^2 M^{-2}$ and dimension of $\left[ B \right]$ is
$L^2$, so
\begin{eqnarray}
A = \left(\frac{c}{\tilde H m}\right)^2, \qquad B =
\left(\frac{c}{\tilde H}\right)^2.
\nonumber
\end{eqnarray}

The interval (\ref{ds2_1}) has now the form
\begin{equation}
\left(\frac{\tilde H}{c}\right)^2ds^2(r,t) =
-e^{\lambda(r,t)}dr^2 - e^{\omega(r,t)}
\left(
d\theta^2 + sin^2\theta d\phi
\right) +d\tau^2,
\label{ds2_D-less}
\end{equation}
Together with metrical functions $\omega(\mu,\tau)$ and $\lambda(\mu,\tau)$,
introduced by Tolman, it is conveniently to use
the Bonnor's function \cite{Bonnor:72}
\begin{equation}
R(\mu,\tau) = e^{\omega(\mu,\tau)/2}.
\label{Bonnor}
\end{equation}
In the Bonnor's notation the interval (\ref{ds2_D-less}) takes the form
\begin{eqnarray}
\left(\frac{\tilde H}{c}\right)^2 ds^2(\mu,\tau) =
-\frac{\left[R^{\prime}(\mu,\tau)\right]^2}{f^2} d\mu^2 -
R^2(\mu,\tau)
\left(
d\theta^2 + sin^2\theta d\phi
\right) + d\tau^2
\nonumber
\end{eqnarray}
As it is shown in \cite{L&L} and \cite{LE&AD}, the Bonnor's coordinate
$R(\mu,\tau)$ has
a sense of Euler coordinate, so the equation (\ref{Bonnor})
correlates geometrical radius of the sphere $R(\mu,\tau)$ where the
particle is located, and the Lagrangian coordinate $\mu$ of this sphere.

To describe the radial motion we will use the Habble function connected
with variation of the radial length $d l$:
\begin{equation}
h = \frac{d \dot l}{d l},
\label{habble_def}
\end{equation}
where, according to the (\ref{ds2_D-less}), for $dl^2$ we read:
\begin{equation}
dl^2 = e^{\lambda(\mu,\tau)} d \mu^2  =
\left(
\frac{R^{\prime}(\mu,\tau)}{f(\mu)} d\mu
\right)^2.
\label{d l}
\end{equation}
By the substitution (\ref{d l}) into the definition of the Habble
function (\ref{habble_def}) we obtain
\begin{equation}
h(\mu,\tau) = \frac{\dot \lambda(\mu,\tau)}{2} = \frac{\dot R
^{\prime}
(\mu,\tau)
}{R^{\prime}(\mu,\tau)}
= \frac{\partial \ln{R^{\prime}(\mu,\tau)}}{\partial \tau}
\label{Habble_def}
\end{equation}
By the integration of the equation (\ref{Habble_def})
we obtain the formulum for metrical function $\lambda(\mu,\tau)$:
\begin{eqnarray}
\lambda(\mu,\tau) = 2 \int\limits_{0}^{\tau}h(\mu,\tau)d\tau +
\lambda(\mu,0)
\nonumber
\end{eqnarray}

A solution in the LTB model is defined by the
functions $f(r)$, $\it F(r)$ and $\bf F\it(r)$.
These functions are obtained in the process of solution of the system of
PDE, so to define them the initial/boundary conditions should
definitely be used.
The metrical function $\omega(\mu,\tau)$ is the solution of the equation
(\ref{T:12}).
The equations of the LTB model are obtained in \cite{Tolman:34}
and solved in the parametric form
for the three cases $f^2(\mu) < 1$, $f^2(\mu) = 1$ and $f^2(\mu) >
1$ in \cite{Bonnor:72} and \cite{Bonnor:74}.

The equations (\ref{T:10}) - (\ref{T:12}) are valid for every $\tau$,
and due to this fact in the Cauchy problem they {\it definde} the functions
$f^2(\mu)$, $\it F(\mu)$ and $\bf F\it(\mu)$ at the moment of time
$\tau = 0$:
\begin{equation}
{\rm
f^2(\mu) - 1 =
e^{\omega_0(\mu)}
\left(
\ddot \omega_0(\mu) + \frac{3}{4} \dot \omega^2_0(\mu) - \Lambda
\right)
}
\label{DEF:f^2-1}
\end{equation}
\begin{equation}
F{\rm(\mu)} =
{\rm
e^{3 \,\omega_0(\mu) / 2}
\left(
\frac{\dot \omega^2_0(\mu)}{2} - \frac{2}{3} \Lambda
\right)
+2
e^{\omega_0(\mu) /2}
\left[1 - f^2(\mu)\right]}
\label{DEF:itF}
\end{equation}
\begin{equation}
{\bf F} {\rm (\mu)} =
{\rm
\int \limits_{
e^{\omega_0(0)/2}
}^{
e^{\omega_0(\mu)/2}
} }
\frac{
d x
      }
     {
      \sqrt{
f^2(\mu) - 1 +\frac{F(\mu)}{2 x}
%
+
            \frac{\Lambda}{3}
x^2
           }
     }
\label{DEF:F}
\end{equation}
At the time $\tau = 0$ the equation (\ref{T:8}) {\it defindes} the function
$\lambda_0(\mu)$:
\begin{equation}
e^{\lambda_0(\mu)} =
e^{\omega_0(\mu)}
\frac{[\omega_0(\mu)]^{\prime \,2}}{4f^2(\mu)}.
\label{DEF:lambda_0_new)}
\end{equation}
Substituting (\ref{DEF:f^2-1}) into (\ref{DEF:itF}), we obtain
\begin{equation}
F
{\rm
(\mu) = e^{3\omega_0(\mu)/2}\left(
-2 \ddot\omega_0(\mu) - \dot\omega^2_0(\mu) +
\frac{4}{3}\Lambda
\right)
}.
\label{DEF:itF_new}
\end{equation}

Comparing (\ref{T:8}) - (\ref{T:11}) with (\ref{DEF:f^2-1}) -
(\ref{DEF:itF}), we find out that
%
\begin{eqnarray}
f^2(\mu) - 1 =
e^{\omega_0(\mu)}
\left(
\ddot \omega_0(\mu) + \frac{3}{4} \dot \omega^2_0(\mu) - \Lambda
\right) &=& \nonumber\\
e^{\omega(\mu,\tau)}
\left(
\ddot \omega(\mu,\tau) + \frac{3}{4} \dot \omega^2(\mu,\tau) - \Lambda
\right)
\label{DEF:f^2-1_n}
\end{eqnarray}
and
\begin{eqnarray}
F(\mu) =
e^{3\omega_0(\mu)/2}\left(
-2 \ddot\omega_0(\mu) - \dot\omega^2_0(\mu) +
\frac{4}{3}\Lambda
\right) &=& \nonumber\\
e^{3\omega(\mu,\tau)/2}\left(
-2 \ddot\omega(\mu,\tau) - \dot\omega^2(\mu,\tau) +
\frac{4}{3}\Lambda
\right)
\label{DEF:itF_new_n}
\end{eqnarray}
are not dependent on time.
Let's use the previous results to calculate the functions ${\bf F}(\mu)$
and integral in the equation (\ref{T:12}).
Substituting the definitions (\ref{DEF:f^2-1}) and (\ref{DEF:itF_new}) into
(\ref{DEF:F}) we obtain:
\begin{equation}
{\bf F}(\mu) =
\pm \int \limits_{\omega_0(0)}^{\omega_0(\mu)}
\frac{d \tilde\omega}{\dot{\tilde\omega}}.
\nonumber
\end{equation}

The function ${\bf F}(\mu)$ is equal to zero at the moment of time
$\tau = 0$ according the definition.
Substituting the right part of the equations (\ref{DEF:f^2-1_n}) and
(\ref{DEF:itF_new_n}) into the (\ref{DEF:F}), we obtain the equation
\begin{equation}
\pm\int \limits^{\omega(\mu,\tau)}_{\omega(\mu,0)}
\frac{d \tilde\omega}{\dot{\tilde\omega}} =
\pm \int \limits_{\omega_0(0)}^{\omega_0(\mu)}
\frac{d \tilde\omega}{\dot{\tilde\omega}} + \tau.
\nonumber
\end{equation}

This analysis of the LTB model shows that the functions
\begin{equation}
\left.
\begin{array}{c}
\left.\omega(\mu,\tau)\right|_{\tau=0} = \omega_0(\mu) \quad
\left.\dot\omega(\mu,\tau)\right|_{\tau=0} = \dot\omega_0(\mu) \quad \\ \\
\left.\ddot\omega(\mu,\tau)\right|_{\tau=0} = \ddot\omega_0(\mu),
\end{array}
\right\}
\label{init}
\end{equation}
and constants
\begin{equation}
\left.
\begin{array}{c}
\left.\omega(\mu,0)\right|_{\mu=0} = \omega_0(0) \quad
\left.\dot\omega(\mu,0)\right|_{\mu=0} = \dot\omega_0(0) \quad \\ \\
\left.\ddot\omega(\mu,0)\right|_{\mu=0} = \ddot\omega_0(0), \quad
\Lambda,
\end{array}
\right\}
\label{init-c}
\end{equation}
are included into the definitions (\ref{DEF:f^2-1}) - (\ref{DEF:F})
and they form the initial conditions of the Cauchy problem for the
equations (\ref{T:4}) - (\ref{T:7}).
In accordance with (\ref{DEF:lambda_0_new)})
the function $\lambda_0(\mu)$ is not include in
the set of initial conditions.

Substituting (\ref{DEF:itF_new}) into the (\ref{T:15}),
we obtain the general expression for the density of "gas" in the LTB model:
\begin{eqnarray}
{\rm
    8 \pi \delta(\mu,\tau) =
    \frac{e^{
             \frac{3}{2}
             [\omega_0(\mu) - \omega(\mu,\tau)]
            }
         }
         {
          \omega^{\prime}(\mu,\tau)
         }
} \times
\nonumber\\
{\rm
    \left\{
           3\left[\omega_0(\mu)\right]^{\prime}
            \left[
                  -\ddot\omega_0(\mu) - \frac{1}{2}\dot\omega_0^2(\mu) +
                  \frac{\Lambda}{6}
            \right]
     - 2\left[\ddot\omega_0(\mu)\right]^{\prime} -
     2\dot\omega_0(\mu)\left[\dot\omega_0(\mu)\right]^{\prime}
    \right\}
}
\label{DEF:rho_new}
\end{eqnarray}
We obtain the formulum for Habble's function using (\ref{T:12}),
(\ref{Bonnor}) and (\ref{Habble_def}):
\begin{equation}
h(\mu,\tau) =
\frac{
      \frac{\partial K(\mu,\tau)}{\partial \mu}
     }
     {
      \frac{\partial}{\partial \mu}\int\limits_{0}^{\tau} K(\mu,\tau)d \tau
      +R^{\prime}(\mu,0)
     },
\label{Habbl-ini}
\end{equation}
where
\begin{eqnarray}
K(\mu,\tau) = \sqrt{f^2(\mu) - 1 + \frac{F(\mu)}{2 R(\mu,\tau)} +
\frac{\Lambda}{3}R^3(\mu,\tau)}.
\nonumber
\end{eqnarray}
The dependence of cosmologycal parameter on the initial
conditions of the LTB model
is represented by the follow formula:
\begin{eqnarray}
\Omega(\mu,\tau) = \frac{\delta(\mu,\tau)}{\delta_c(\mu,\tau)},
\nonumber
\end{eqnarray}
where the critical density defined by formulum
\begin{eqnarray}
\delta_{c}(\mu,\tau) = \frac{3 \tilde{H}^2}{8 \pi G \rho_0} h^2(\mu,\tau).
\nonumber
\end{eqnarray}
The function $\omega(\mu,\tau)$ from the equation (\ref{DEF:rho_new})
is the solution of the equation (\ref{T:12}).

\section{Results}

The LTB model is the Cauchy problem for the PDE (\ref{T:4}) - (\ref{T:7}).
Traditionally, three functions $f(\mu)$, $F(\mu)$ and ${\bf F}(\mu)$
are used to chose some physical propetrys of the problem solving in the LTB
model. These functions play a role of the initial conditions in the
LTB model. But the input
equations (\ref{T:4}) - (\ref{T:7}) are wrtitten throught the metrical
functions
$\lambda(\mu,\tau)$ and $\omega(\mu,\tau)$. So, should definitly be studied
the dependence functions $f(\mu)$, $F(\mu)$ and ${\bf F}(\mu)$ on
metrical functions. The present article learns this
rpoblem and three initial conditions (\ref{init}) - (\ref{init-c})
defined three function
by the rules (\ref{DEF:f^2-1}) - (\ref{DEF:F}).
The general expressions for the density (\ref{DEF:rho_new}) and Habble
fuction (\ref{Habbl-ini}) show the dependence on the initial conditions
(\ref{init}).

\section{Acknowledgements}

I'm grateful to Prof. Arthur D. Chernin for encouragement and discussion.
Dr. Yurij Barishev has initiated my interest to the modern Cosmology as
fractal structure of the Universe and interpretation of the observations by
calculation redshift in the LTB model.
This paper was financially supported by "COSMION" Ltd., Moscow.

\end{document}